\documentclass[11pt,a4paper]{article}
\usepackage{jheppub}
\pdfoutput=1
\usepackage[english]{babel}
\usepackage[latin1]{inputenc} 

\usepackage{color}
\usepackage{float}
\usepackage{amsxtra}
\usepackage{amstext}
\usepackage{amssymb}
\usepackage{latexsym}
\usepackage{dsfont}
\usepackage{hyperref}

\usepackage{multirow}
\usepackage{booktabs}
\usepackage{ifthen}

\usepackage{rotating}
\usepackage{longtable}
\usepackage{pdflscape}

\usepackage{appendix}

\newcommand{\forloop}[5][1]{%
\setcounter{#2}{#3}%
\ifthenelse{#4}{#5\addtocounter{#2}{#1}%
\forloop[#1]{#2}{\value{#2}}{#4}{#5}}%
{}}

\newcounter{crcounter}

\newcommand{\compensaterule}[1]{%
\forloop{crcounter}{1}{\value{crcounter} < #1}%
{\vspace*{-\aboverulesep}\vspace*{-\belowrulesep}}}

\newcommand{\multirowbt}[4][\anzlines]{%
\def\anzlines{#2}%
\multirow{#2}{#3}%
{\compensaterule{#1}#4}}

\newcommand{\lumipb}{pb$^{-1}$}
\newcommand{\lumifb}{fb$^{-1}$}
\newcommand{\intlumi}{$L_{\text{int}}$}

\newcommand{\lumieins}{$L_{\text{int}}=1.04$ fb$^{-1}$}
\newcommand{\lumitext}{integrated luminosity}
\newcommand{\gev}{GeV}
\newcommand{\gevklam}{[GeV]}
\newcommand{\etmis}{E$_{\text{T}}^{\text{miss}}$}
\newcommand{\meff}{m$_{\text{eff}}$}
\newcommand{\pt}{p$_{\text{T}}$}

\newcommand{\atl}{ATLAS}

\newcommand{\zerolepton}{``Search for squarks and gluinos using final states with jets and missing transverse momentum''}
\newcommand{\modno}{20000}

\title{Constraints on the pMSSM 
from searches for squarks and gluinos by ATLAS}
\author[a]{Antonia Str\"ubig,}
\author[b]{Sascha Caron}
\author[a]{and Michael Rammensee}

\affiliation[a]{Institute for Physics, Albert-Ludwigs-Universität Freiburg, Hermann-Herder-Str.3, 79109 Freiburg, Germany}
\affiliation[b]{Institute for Mathematics, Astrophysics and Particle Physics, Radboud University Nijmegen, Nijmegen and NIKHEF, Science Park, Amsterdam, The Netherlands}

\emailAdd{antonia.struebig@cern.ch}
\emailAdd{scaron@cern.ch}
\emailAdd{michael.christian.rammensee@cern.ch}

\abstract{
We study the impact of the jets and missing transverse momentum SUSY analyses  of the ATLAS experiment on the phenomenological MSSM (pMSSM).
We investigate sets of SUSY models with a flat and logarithmic prior 
in the SUSY mass scale and a mass range up to 1 and 3 TeV, respectively. 
These models were found previously in the study 'Supersymmetry without Prejudice'. 
Removing models with long-lived SUSY particles, we show that 
99\% of $20000$ randomly generated pMSSM model points with a flat prior
 and 87\% for a logarithmic prior are excluded by the ATLAS results. 
For models with squarks 
and gluinos below $600$~ GeV all models of the pMSSM grid are excluded.
We identify SUSY spectra where the current ATLAS search strategy is less sensitive and propose extensions to the inclusive jets search channel. 
}


\begin{document}
\maketitle

\section{Introduction}
Supersymmetry\cite{supersym} is one of the conceivable extensions of the Standard Model (SM). It could provide a natural candidate for cold dark matter and stabilise the electroweak scale by reducing the fine tuning of higher order corrections to the Higgs mass. Supersymmetry (SUSY) proposes superpartners for the existing particles. Squarks and gluinos, superpartners of the quarks and the gluon are heavy coloured particles, which can decay to jets and the Lightest Supersymmetric Particle (LSP), i.e. the neutralino. The neutralino is only weakly interacting and stable since we assume the conservation of R-parity.
The LSP escapes detection 
which results in missing transverse momentum in the detector. 
Channels with jets and missing transverse momentum have a large discovery potential at 
the LHC \cite{AtlasExp}, since the coupling strength of the strong force would cause an 
abundance of squarks and gluinos if these particles are not too heavy. 
The ATLAS collaboration has analysed their data to search for squarks and gluinos 
in events with 2-4 jets and missing transverse momentum  corresponding to an integrated 
luminosity \intlumi\ of 35 \lumipb\ in Ref.~\cite{0lep0211} and 1.04 \lumifb in Ref.~\cite{0lep0911}. 
No excess above the SM background expectation was observed in the analysed data. Although these searches are designed to be quite independent of SUSY model assumptions, mass limits are presented only for a constrained Minimal Supersymmetry Standard Model (cMSSM) model and for simplified models with only squarks, gluinos and the lightest neutralino.\\

We will study the exclusion range of the \atl\ search for phenomenological MSSM (pMSSM)\cite{pmssm} scenarios, which have a more diverse spectrum of characteristics than the cMSSM. We identify some of the
regions in the pMSSM parameter space where the current search strategy is 
insensitive.

In the pMSSM the more than 120 free parameters of 
the MSSM are reduced to 19 by demanding CP-conservation, minimal flavor violation and degenerate mass spectra for the 
1st and 2nd generations of sfermions. 
In addition it is required that the LSP is the neutralino $\tilde{\chi}_1^0$. The 19 remaining parameters are 
10 sfermion masses\footnote{The sfermion parameters are 
\ensuremath{\tilde{Q}_{\mathrm{L}}}, 
\ensuremath{\tilde{Q}_{\mathrm{3}}}, 
\ensuremath{\tilde{L}_{\mathrm{1}}}, 
\ensuremath{\tilde{L}_{\mathrm{3}}}, 
\ensuremath{\tilde{u}_{\mathrm{1}}}, 
\ensuremath{\tilde{d}_{\mathrm{1}}}, 
\ensuremath{\tilde{u}_{\mathrm{3}}}, 
\ensuremath{\tilde{d}_{\mathrm{3}}}, 
\ensuremath{\tilde{e}_{\mathrm{1}}} and 
\ensuremath{\tilde{e}_{\mathrm{3}}}. 
},  3 gaugino masses $M_{1,2,3}$, the ratio of the Higgs vacuum expectation values tan $\beta$, the Higgsino mixing parameter $\mu$, the pseudoscalar Higgs boson mass $m_A$ and 3 A-terms $A_{b,t,\tau}$. This work is based on ``Supersymmetry Without Prejudice''\cite{susynoprejudice}. The model points presented in \cite{susynoprejudice} are used for our purpose. Each model point was constructed by a quasi-random sampling of the pMSSM parameters 
space. The points were required to be consistent 
with the experimental constraints prior to the LHC ~\cite{susynoprejudice}.

\section{Event generation, Fast Simulation and Analysis}
We study the reach of the ATLAS search by emulating the ATLAS analysis chain. First we generate
events from LHC collisions for each pMSSM SUSY model with a Monte Carlo generator for SUSY processes. 
These events are then simulated by a fast detector simulation
and the acceptance and efficiency is determined by applying the most important ATLAS analysis cuts on the simulated events. Finally these numbers
are used to calculate the expected number of signal events for each signal region and analysis. These numbers are compared to the 
model-independent $95\%$ C.L. limits provided by ATLAS.  

PYTHIA 6.4\cite{pythia} is used for the event simulation of proton-proton
collisions at a 7~TeV centre-of-mass energy.
All squark and gluino production processes 
are enabled as they are of most importance 
for the inclusive jets search channel. For every 
model point 10000 events are generated which we found to be enough even for the models with the largest cross sections. 
To get as close as possible to the ATLAS analysis we use DELPHES 1.9\cite{delphes} as a fast detector 
simulation with the default ATLAS detector card, modified by setting the jet cone radius to $0.4$. 
The PYTHIA output is read in by DELPHES in HepMC format, which is produced by 
HepMC 2.04.02\cite{hepmc}. The object reconstruction is done by DELPHES, which uses the same anti-k$_T$ jet algorithm\cite{antikt} as ATLAS. Also included in the reconstruction are isolation criteria
for electrons and muons. We do not emulate pile-up events.

Reconstructed events are analysed with the same event selections as used by the \atl\ analysis with 35 \lumipb\ 
(shown in Table~\ref{tab:sigreg35}) and also with the event selections used in the $1.04$ \lumifb\ analysis (see Table~\ref{tab:sigreg104}). In these Tables $\varDelta\phi(\text{jet}_i,\text{E}_{\text{T}}^{\text{miss}})_{min}$ is the minimum of the azimuthal angles between the jets and the 2-vector of the missing transverse momentum $\vec{E}_T^{\text{miss}}$. The invariant mass \meff\ is calculated as the scalar sum of \etmis\ and the magnitudes of the \pt\ of the leading jets required in the selection (i.e. 2 jets for the 2-jet selection in region A), except for signal region E, where \meff\ is the sum of \etmis\ and all reconstructed jets with p$_\text{T}>40$ GeV. In addition to these cuts a veto on electrons and muons with \pt$>20$ GeV was required.\\
 
After this selection the event counts are scaled to the luminosities considered in the analyses, i.e. 35 \lumipb\ and 1.04 \lumifb, respectively. The NLO cross section used for this is calculated by LHC-Faser light\cite{LHCFaserweb,LHCFaserVor} from PROSPINO2.1~\cite{prospino1,prospino2} cross section grids.\\
The limits on the effective cross sections given by the \atl\ analyses are used to calculate a limit on the number of signal events passing the cuts, also given in Table~\ref{tab:sigreg35} and \ref{tab:sigreg104}.
No attempt was made to include theoretical uncertainties. In the studied SUSY mass range these uncertainties are small compared to the differences of the ATLAS and DELPHES setups and would not change drastically any conclusion of this work.   

\begin{table}[h]
\begin{center}
\begin{tabular}{cccc}
\toprule
Signal region: & A & C & D\\
\midrule
\etmis\ \gevklam& \multicolumn{3}{c}{\begin{footnotesize} for all regions \end{footnotesize} $>$100}\\
leading jet \pt\ \gevklam& \multicolumn{3}{c}{\begin{footnotesize} for all regions \end{footnotesize} $>$120}\\
2nd jet \pt\ \gevklam& $>$40 & $>$40 & $>$40\\
3rd jet \pt\ \gevklam& - & $>$40 & $>$40\\
$\varDelta\phi(\text{jet}_i,\text{E}_{\text{T}}^{\text{miss}})_{min}$& \multicolumn{3}{c}{\begin{footnotesize} for all regions \end{footnotesize}$>$0.4}\\
\meff\ \gevklam & $>$500& $>$500& $>$1000\\
f=\etmis/\meff &  $>$0.3& $>$0.25& $>$0.25\\
\midrule
$95\%$ C.L. limit on $\sigma$ [pb] & 1.3 & 1.1 & 0.11 \\ 
\bottomrule
\end{tabular}
\end{center}
\caption{Requirements for the signal regions A,C and D for the ATLAS analysis with an \lumitext\ of 35 \lumipb. In addition the number of reconstructed leptons has to be zero; also shown are the $95\%$ C.L. upper limits on the cross section for new physics processes $\sigma$.}
\label{tab:sigreg35}
\end{table}

\begin{table}[h]
\begin{center}
\begin{tabular}{cccccc}
\toprule
Signal region: & A & B & C & D & E\\
\midrule
\etmis\ \gevklam& \multicolumn{5}{c}{\begin{footnotesize} for all regions \end{footnotesize} $>$130}\\
leading jet \pt\ \gev& \multicolumn{5}{c}{\begin{footnotesize} for all regions \end{footnotesize} $>$130}\\
2nd jet \pt\ \gevklam& $>$40& $>$40& $>$40& $>$40& $>$80\\
3rd jet \pt\ \gevklam&  -  & $>$40& $>$40& $>$40& $>$80\\
4th jet \pt\ \gevklam&  -  &  -  & $>$40& $>$40& $>$80\\
$\varDelta\phi(\text{jet}_i,\text{E}_{\text{T}}^{\text{miss}})_{min}$& \multicolumn{5}{c}{\begin{footnotesize} for all regions \end{footnotesize} $>$0.4}\\
\meff\ \gev & $>$1000& $>$1000& $>$500& $>$1000& $>$1100\\
f=\etmis/\meff &  $>$0.3& $>$0.25& $>$0.25& $>$0.25& $>$0.3\\
\midrule
$95 \%$ C.L. limit on $\sigma$ [fb] & 22 & 25 & 429 & 27 & 17\\ 
\bottomrule
\end{tabular}
\end{center}
\caption{Requirements for the signal regions A - E for the ATLAS analysis with an \lumitext\ of 1.04 \lumipb. In addition the number of reconstructed leptons has to be zero; also shown are the $95\%$ C.L. upper limits on the cross section for new physics processes $\sigma$.}
\label{tab:sigreg104}
\end{table}

In order to compare our setup to ATLAS we determined
the relative efficiency difference
\begin{displaymath}
\frac {\Delta C} {C} = \frac{(A*E)_{\rm{ATLAS}} - (A*E)_{\rm{DELPHES}}} {(A*E)_{\rm{ATLAS}}} 
\end{displaymath}

for each SUSY point studied by ATLAS 
in the $m_0$-$m_{1/2}$ plane for the cMSSM grid with $\tan \beta = 10 $, $A_0=0$ and $\mu>0$. 
Here $A*E$ is the acceptance times efficiency of the ATLAS and DELPHES analysis setups.  

\begin{table}[h]
\begin{center}
\begin{scriptsize}
\begin{tabular}{lllccccc}
\toprule
\multirow{2}{*}{m$_0$}&\multirow{2}{*}{m$_{1/2}$}&&\multicolumn{5}{c}{Accepted fraction of signal events per signal region}\\
& &	                                          & A & B & C & D & E\\
\midrule
\multirowbt[4]{6}{*}{340}& \multirowbt{2}{*}{120}& {\scriptsize ATLAS}& 0.001 & 0.002 & 0.08 & 0.002 & 0.003 \\
&&        {\scriptsize DELPHES}&  0.002$\pm$0.0004 & 0.003$\pm$0.0005 & 0.06$\pm$0.003 & 0.003$\pm$0.0005 & 0.004$\pm$0.0006\\
\cmidrule{4 - 8}
&    \multirowbt{2}{*}{300}& {\scriptsize ATLAS}& 0.1 & 0.13 & 0.19 & 0.11 & 0.09\\
&&        {\scriptsize DELPHES}& 0.15$\pm$0.004 & 0.14$\pm$0.004 & 0.16$\pm$0.004 & 0.1$\pm$0.003 & 0.06$\pm$0.003\\
\cmidrule{4 - 8}
&    \multirowbt{2}{*}{450}& {\scriptsize ATLAS}& 0.27 & 0.26 & 0.23 & 0.2 & 0.15\\
&&        {\scriptsize DELPHES}& 0.33$\pm$0.006 & 0.27$\pm$0.005 & 0.18$\pm$0.004 & 0.17$\pm$0.004 & 0.11$\pm$0.003\\
\midrule
\multirowbt[4]{6}{*}{1140}& \multirowbt{2}{*}{120}& {\scriptsize ATLAS}& 0.002 & 0.003 &  0.08 & 0.004 & 0.004\\
&&        {\scriptsize DELPHES}& 0.003$\pm$0.0006 & 0.004$\pm$0.0006 & 0.06$\pm$0.002 & 0.004$\pm$0.0006 & 0.003$\pm$0.0005\\
\cmidrule{4 - 8}
&    \multirowbt{2}{*}{300}& {\scriptsize ATLAS}& 0.05 & 0.07 & 0.13 & 0.08 & 0.14\\
&&        {\scriptsize DELPHES}& 0.05$\pm$0.002 & 0.07$\pm$0.003 & 0.1$\pm$0.003 & 0.07$\pm$0.003 & 0.09$\pm$0.003\\
\cmidrule{4 - 8}
&    \multirowbt{2}{*}{450}& {\scriptsize ATLAS}& 0.12 & 0.16 & 0.18 & 0.16 & 0.2\\
&&        {\scriptsize DELPHES}& 0.09$\pm$0.003 & 0.09$\pm$0.003 & 0.08$\pm$0.003 & 0.08$\pm$0.003 & 0.1$\pm$0.003\\
\midrule
\multirowbt[4]{6}{*}{2500}& \multirowbt{2}{*}{120}&{\scriptsize ATLAS}& 0.0001 & 0.002 &  0.07 & 0.002 & 0.003 \\ 
&&        {\scriptsize DELPHES}& 0.001$\pm$0.0003 & 0.002$\pm$0.0004 & 0.07$\pm$0.003 & 0.002 $\pm$0.0004 & 0.003$\pm$0.0005\\
\cmidrule{4 - 8}
&    \multirowbt{2}{*}{300}& {\scriptsize ATLAS}& 0.02 & 0.05 & 0.12 & 0.07 & 0.11\\
&&        {\scriptsize DELPHES}& 0.02$\pm$0.001 & 0.04$\pm$0.002 & 0.08$\pm$0.003 & 0.04$\pm$0.002 & 0.07$\pm$0.003\\
\cmidrule{4 - 8}
&    \multirowbt{2}{*}{360}& {\scriptsize ATLAS}& 0.03 & 0.07 & 0.13 & 0.08 & 0.15\\
&&        {\scriptsize DELPHES}& 0.03$\pm$0.002 & 0.04$\pm$0.002 & 0.07$\pm$0.003 & 0.05$\pm$0.002 & 0.08$\pm$0.003\\
\bottomrule
\end{tabular}
\end{scriptsize}
\end{center}
\caption{Accepted signal fraction ($E*A$) for the ATLAS and DELPHES setup and shown for the analysis
with \lumieins.}
\label{tab:effEPS}  
\end{table}

\begin{figure}[h]
  \centering
  \includegraphics[width=0.45\textwidth]{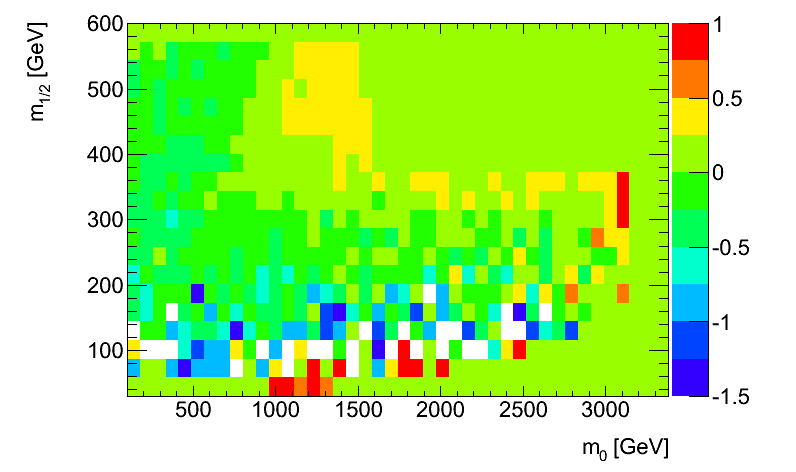}
  \includegraphics[width=0.45\textwidth]{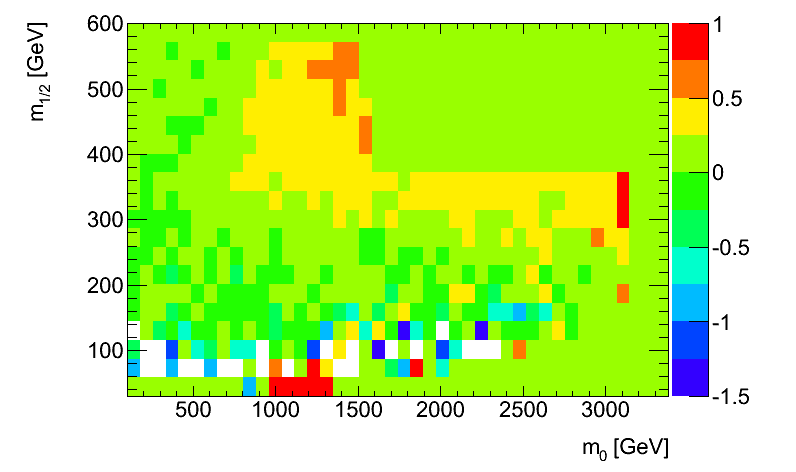}
  \includegraphics[width=0.45\textwidth]{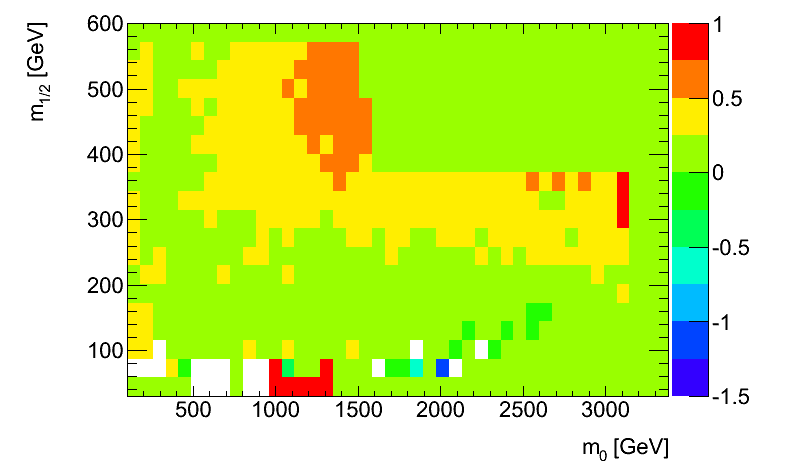}
  \includegraphics[width=0.45\textwidth]{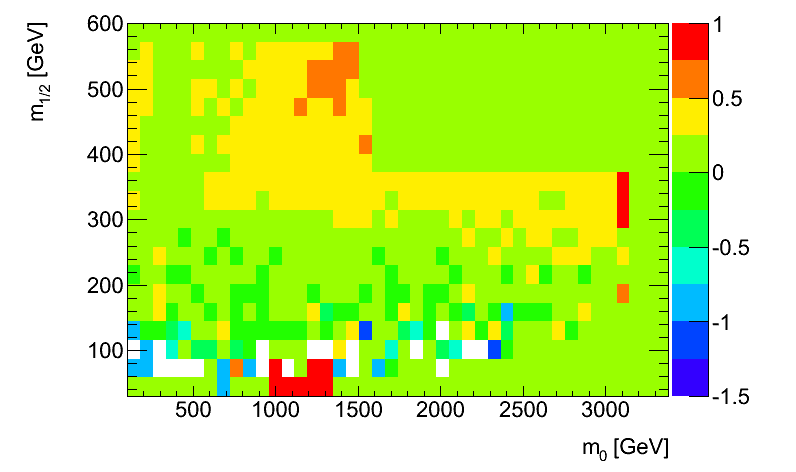}
  \includegraphics[width=0.45\textwidth]{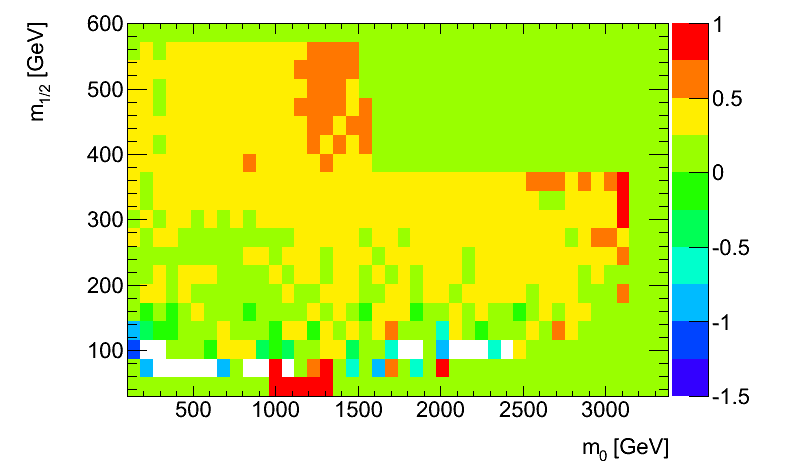}
 \caption{Relative efficiency times acceptance difference $\frac{\Delta C}{C}$ of the ATLAS and DELPHES analysis setups for signal region $A-E$ (top left to bottom right) in the m$_0$-m$_{1/2}$ plane of the cMSSM model studied by ATLAS. The difference exceeds the indicated range in the white areas. ATLAS has not provided numbers for the upper right box, here the differences was set to 0.}
  \label{fig:eff}
\end{figure}

Figure~\ref{fig:eff} shows $\frac{\Delta C} {C}$ for the cMSSM.
Numerical examples for the ATLAS and DELPHES efficiencies 
are shown in Table~\ref{tab:effEPS}.   
The efficiency of our setup is found to be in agreement with 
the ATLAS efficiency\cite{hepdata} on the level of $10-30\%$
for the 2- and 3-jet signal regions $A-B$ and SUSY masses around the present 
ATLAS limits. These limits are ranging 
for $m_{1/2}$ from $200-500$~GeV and go up to intermediate $m_{0}$ of $1000$~GeV.

At $m_{1/2}<200$~GeV larger deviations occur. Here both the statistical uncertainty of
the ATLAS and DELPHES efficiencies 
are larger  and the selection efficiencies are tiny. The largest deviations occur 
if in addition $m_{0}$ is large.  
The signal regions are not intended for 
SUSY signals at $m_{1/2}<200$~GeV and large $m_0$ and do therefore not
contribute to the search for such SUSY signals. 
Note that the ATLAS analysis selects always the signal region with the
largest exclusion potential for each SUSY model.

For signal region $C-E$ and for the 4 and more jet channels we observe better
agreement at low $m_{1/2}$ and slightly worse agreement
at $m_0>1000$~GeV and $m_{1/2}>400$~GeV. Here 
our DELPHES setup underestimates the efficiency by
up to $50-70\%$.  
The increased differences at larger jet multiplicities  
might be caused by the cummulative effects of the
the ATLAS and DELPHES jet response.

In view of the mostly 
smaller efficiencies of DELPHES compared to ATLAS, our study can be regarded as conservative.

\section{pMSSM random points}
The pMSSM points are taken from ``Supersymmetry Without Prejudice''\cite{susynoprejudice} (related work \cite{nopredLHC,nopred7tev}). All
details can be found in these references. 
19 free parameters were randomly sampled, one set with a 
flat prior with masses up to 1 TeV and another 
one with a logarithmic prior and masses up to 3 TeV, each parameter was 
varied in the range given in Table~\ref{tab:parrange}. The parameters 
are: 10 sfermion masses $m_{\tilde{f}}$; 3 gaugino masses $M_{1,2,3}$; the ratio of the Higgs vacuum expectation values tan $\beta$; the Higgsino mixing parameter $\mu$; the pseudoscalar Higgs boson mass $m_A$ and 3 A-terms $A_{b,t,\tau}$, the A-terms for the first and second generations can be neglected due to the small Yukawa couplings.\\

\begin{table}[h]
\begin{center}
\begin{tabular}{ccc}
\toprule
parameter & flat prior set & log prior set\\
\midrule
$m_{\tilde{f}}$ & 100 GeV - 1 TeV & 100 GeV - 3 TeV\\
$|M_{1,2},\mu|$ & 50 GeV - 1 TeV & 10 GeV - 3 TeV\\
$M_3$ & 100 GeV - 1 TeV & 100 GeV - 3 TeV\\
$|A_{b,t,\tau}|$& 0 - 1 TeV & 10 GeV - 3 TeV\\
tan$\beta$ & 1 - 50 & 1 - 60\\
$m_A$ &  43.5 GeV - 1 TeV & 43.5 GeV - 3TeV\\ 
\bottomrule
\end{tabular}
\end{center}
\caption{Parameter range for flat and log prior model sets}
\label{tab:parrange}
\end{table}

It was assumed that the neutralino is the LSP and that the first two 
squark generations are degenerate. 
Several experimental and theoretical 
constraints\cite{constraints} are applied on the 
generated points, i.e. the current dark matter density and constraints from LEP and Tevatron data.

Additionally we have required that the mass splitting between the chargino 
and the lightest neutralino is  $\varDelta$m $>0.05$ \gev\ with $\varDelta m = m_{Chargino} - m_{\tilde{\chi_0}}$, 
to avoid mishandling by PYTHIA. Small mass splittings make 
charginos stable and PYTHIA yields error messages in the hadronization routines and drops these events. The problem
is avoided by a decay of the chargino before the hadronization routine, i.e. by a sufficiently 
large mass splitting between the chargino and the neutralino.
About 1\% of the remaining model points could not be generated with PYTHIA
due to other compressed mass spectra, i.e. due to very small mass differences between SUSY particles. 
Here mostly the mass difference of the sbottom or 
stop to the neutralino was small.
These compressed mass spectra lead to long lived squarks
 which can not be handled by PYTHIA nor by the detector simulation and causes PYTHIA to stop.
These model points are dropped.
The following studies are therefore not valid if the SUSY model leads to long-lived particles in the spectrum besides the lightest neutralino.

\section{Results and Discussion}
\subsection{Models from a linear prior in the SUSY mass scale}
For each SUSY model signal events were generated.
Each event was analysed after a detector simulation with DELPHES and the number of signal events
was determined for each SUSY model and each of the 8 studied signal regions. 
In the following we call ``excluded models'' SUSY models which produced 
a larger number of signal events than excluded
by the ATLAS model-independent limits in at least 
one of the signal regions studied. 
The model-independent limits 
are listed in Table \ref{tab:sigreg35} and Table \ref{tab:sigreg104}.
Only the SUSY models which yield less signal events in all regions are not excluded by these ATLAS searches.
These models are called ``not excluded models''.

\begin{table}
\begin{center}
\begin{scriptsize}
\begin{tabular}{cccccccccccc}
\toprule
\footnotesize model & \footnotesize m$_{\tilde{q}}$ & \footnotesize m$_{\tilde{g}}$ & \footnotesize m$_{\tilde{\chi}_0}$ & \footnotesize $\sigma_{NLO}$ & \footnotesize \etmis & \footnotesize \meff & \footnotesize N$_{Jets}$ & \footnotesize 1$^{st}$ p$^{\text{jet}}_{\text{T}}$ & \footnotesize 2$^{nd}$ p$^{\text{jet}}_{\text{T}}$  & \footnotesize N$_{Lep}$ & \footnotesize $\varDelta\phi$\\ 
\toprule
1956 & 683.7 & 820.9 & 127.4 & 0.4 & 261.6 & 999.4 & 6.5 & 379.6 & 235.9 & 2.2 & 1.0\\
\midrule
2083 & 672.8 & 979.8 & 504.5 & 0.2 & 245.4 & 740.6 & 4.8 & 249.7 & 155.9 & 0.4 & 1.2\\
\midrule
3226 & 826.5 & 645.7 & 404.3 & 0.7 & 179.8 & 602.3 & 6.9 & 212.4 & 126.3 & 0.7 & 1.0\\
\bottomrule 
\end{tabular}
\caption{Important properties of some not excluded pMSSM models 
out of \modno. 
Shown are the mass of the lightest squark in the 1. and 2. generation 
m$_{\tilde{q}}$; the gluino 
mass, m$_{\tilde{g}}$; the mass of the lightest neutralino m$_{\tilde{\chi}_0}$; 
the NLO cross section $\sigma_{NLO}$; the average values of 
\etmis, \meff, the number of jets N$_{Jets}$ and the number of 
leptons N$_{Lep}$; 1$^{st}$ p$^{\text{jet}}_{\text{T}}$ and 2$^{nd}$ p$^{\text{jet}}_{\text{T}}$ are 
the average of the leading and second highest jet \pt\ and $\varDelta\phi$ is the 
average of the $\varDelta\phi(\text{jet}_i,\text{E}_{\text{T}}^{\text{miss}})_{min}$ variable.}
\label{tab:interestundisc}
\end{scriptsize}
\end{center}
\end{table}

\begin{figure}[h]
  \centering
  \includegraphics[width=0.6\textwidth]{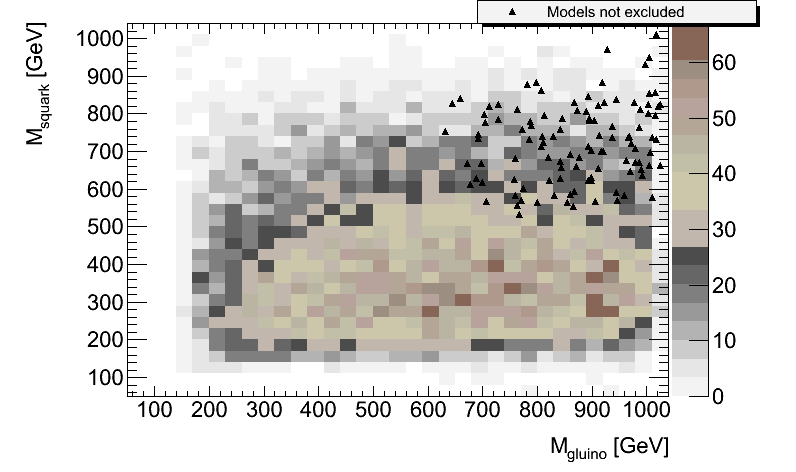}
  \caption{Exclusion range of the ATLAS SUSY analysis for \intlumi\ = 35 \lumipb\ and \intlumi\ = 1.04 \lumifb\ for \modno\ randomly generated pMSSM points with flat prior as a function of the lightest mass of the first and second
generation squarks and the mass of the gluino ; the number of excluded model points for each bin is indicated by the colour scale (bottom figure), not excluded model points are shown in black. }
  \label{fig:excl1035COL}
\end{figure} 

\begin{figure}[h]
  \centering
  \includegraphics[width=0.6\textwidth]{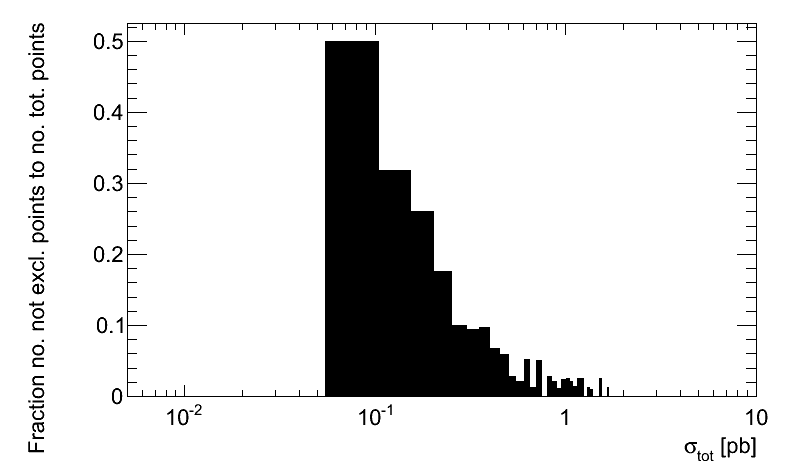}
  \caption{The fraction of the not excluded SUSY models to the total number of studied SUSY models as a function of the NLO cross section for 
squark and gluino production processes.}
  \label{fig:featXsec}
\end{figure}

\begin{figure}[h]
  \centering
  \includegraphics[width=0.6\textwidth]{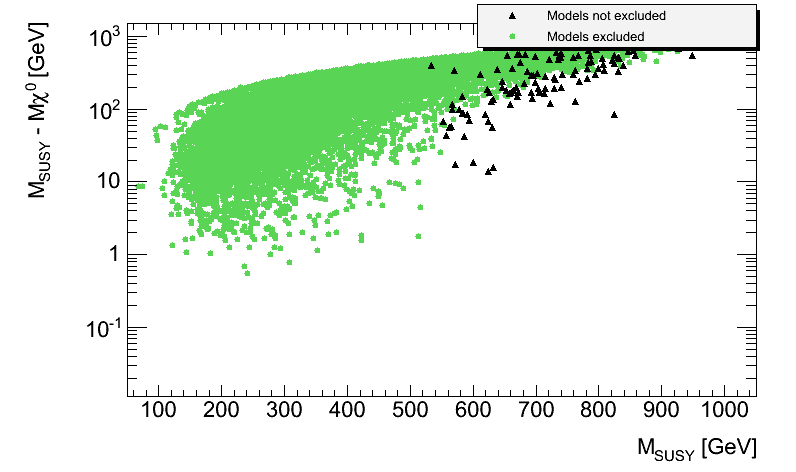}
 \includegraphics[width=0.6\textwidth]{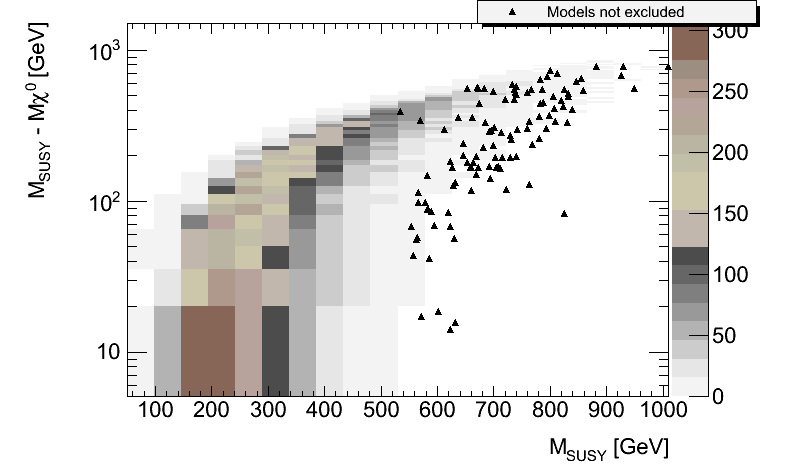}

  \caption{Exclusion range of the ATLAS SUSY analysis for \intlumi\ = 35 \lumipb\ and \intlumi\ = 1.04 \lumifb\ for \modno\ randomly generated pMSSM points with flat prior as a function of $M_{\rm{SUSY}}$ and the mass difference
of $M_{\rm{SUSY}}$ and the mass of the lightest neutralino ; the number of excluded model points for each bin is indicated by the colour scale (bottom figure), not excluded model points are shown in black. The top figure shows the same, but the excluded model points are shown in green.}
  \label{fig:excl1035COLmsusy}
\end{figure}

Figure \ref{fig:excl1035COL} and \ref{fig:excl1035COLmsusy} show the 20000 pMSSM model points from a flat mass prior  as a function of the 
lightest mass of the first and second generation squarks $M_{\rm{squark}}$ and the mass of the gluino $M_{\rm{gluino}}$.
The SUSY 
points which are excluded by ATLAS are shown as green points, models not excluded as black triangles. 
We show that
 99\% of the points are excluded with the current ATLAS analyses in the jets and missing transverse momentum channels. 
All studied points with a mass of the squarks and the mass of the gluino $<600$~GeV are excluded. This means
that there is not much room anymore in the pMSSM 
for having both light squarks of the first generations and at the same time a light gluino. 

Remarkably, also points with small mass splittings between the squarks or gluino and the neutralino are excluded in this mass range.
The reason is quite simple. It is very unlikely that a 
``random'' sampling yields
cases where the mass splittings of {\it all} squarks and the gluino to the neutralino are small. 
If one of the squarks or only the gluino
is a bit heavier than the neutralino
such processes yield detectable rates in the ATLAS signal regions.
Note that in these models the left and right handed squarks can have quite different masses.

In Table~\ref{tab:interestundisc} a subset of the not 
excluded model points are presented together with some of their properties.
A complete list of all not excluded model points can be found in Appendix~\ref{app}.

We found some features why model points are not excluded. 
We determined average values for some properties 
for each SUSY model point, neglecting the fact that these
values are coming from different SUSY decay chains. 
The investigated properties of the non-excluded SUSY models
are shown in Table~\ref{tab:interestundisc}. 
The following features
have been found to be significant:\\

\paragraph{Low cross section}
A large fraction of model points at 
high squark and gluino masses cannot be excluded because the cross section is simply too low to be observed for the \lumitext\ .
Figure~\ref{fig:featXsec} shows the fraction of not-excluded points as a function of the total 
SUSY squark and gluino cross section. Below $0.1$~pb less than $50\%$ of the SUSY models 
can be excluded with the analysis setup. These are mainly 
points with a large average effective mass value.
At large cross sections of greater $5$~pb all studied pMSSM models can be excluded by the ATLAS analyses.

\paragraph{Lepton and multi-jet events (long decay chains)}
 Around 25\% of the not excluded model points have a large average number of leptons. 
In addition we find that these SUSY models do often have a large average number of jets. 
It is trivial to note that, because of the lepton veto, there is not much sensitivity to these models  with the 
inclusive jets analysis. These points can most likely be excluded with the single or multi-lepton analyses. 
These searches do have signal regions investigating events with up to 4 jets~\cite{ATLAS:2011ad,Aad:2011cwa}.
Some SUSY models with long decay chains would yield lepton(s) 
together with multiple jets.  

\paragraph{Compressed spectra together with high squark and gluino masses}
Figure \ref{fig:excl1035COLmsusy} shows the excluded and non-excluded SUSY models  
from the grid with the flat prior as a function of $M_{\rm{SUSY}}$ and the mass difference
of $M_{\rm{SUSY}}$ and the mass of the lightest neutralino. In this note, the SUSY 
mass scale $M_{\rm{SUSY}}$ is defined as the
minimal mass of all first and second generation squarks and the gluino. 
The figure shows the interesting feature that the non-excluded points are mostly located at
small mass differences (relative to $M_{\rm{SUSY}}$) and high $M_{\rm{SUSY}}$.

\begin{figure}[h]

  \centering
\includegraphics[width=0.6\textwidth]{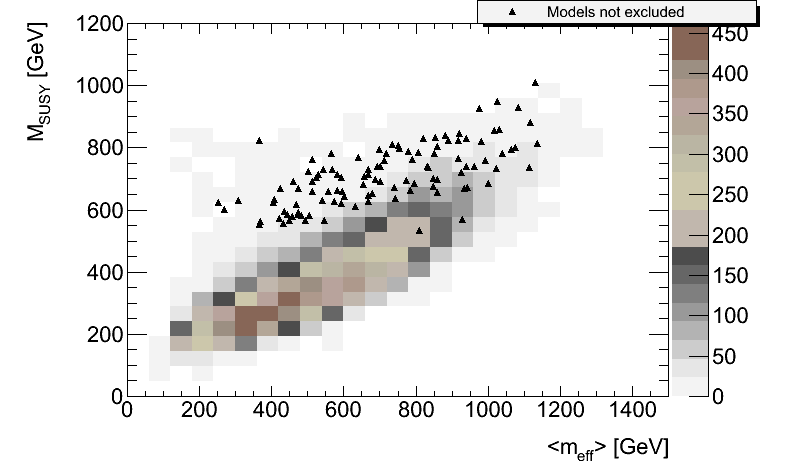}
   \caption{The distribution of m$_{SUSY}$ to the average value of \meff\  for excluded points (colour scale) and not excluded points (black dots).}
   \label{fig:COLmsusy-meff}
\end{figure}

Small mass differences between the colored particles and the neutralino yield
 events with small 
transverse momentum jets.
Figure \ref{fig:COLmsusy-meff} shows the average effective mass (calculated with the leading 3 jets) as a function of $M_{\rm{SUSY}}$. 
More than half of the not-excluded SUSY models at high $M_{\rm{SUSY}}$ have an
effective mass that is significantly below the value found for the excluded SUSY models.
We conclude that the cut on the effective mass is too harsh for these models.
For those compressed 
models the effective mass is differently correlated with the SUSY mass scale.

\begin{figure}[h]
  \centering
\includegraphics[width=0.6\textwidth]{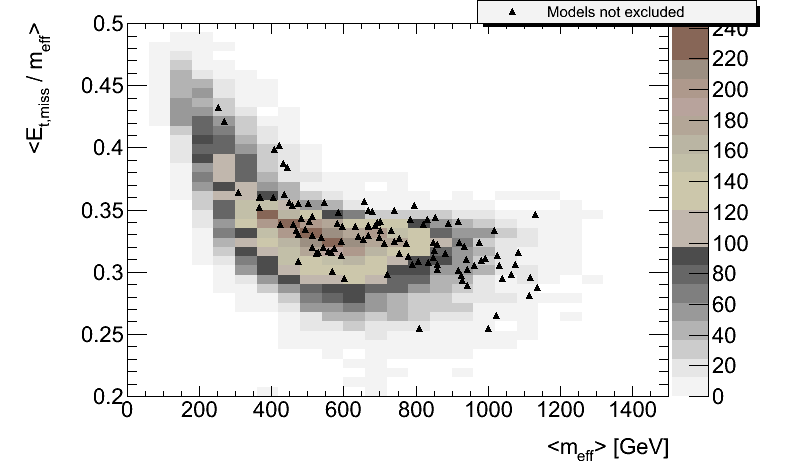}
   \caption{The distribution of  the average value \etmis/\meff\ to the average value of \meff\  for excluded points (colour scale) and not excluded points (black dots).}
   \label{fig:COLettom-meff}
\end{figure}

\begin{figure}[h]
  \centering
  \includegraphics[width=0.6\textwidth]{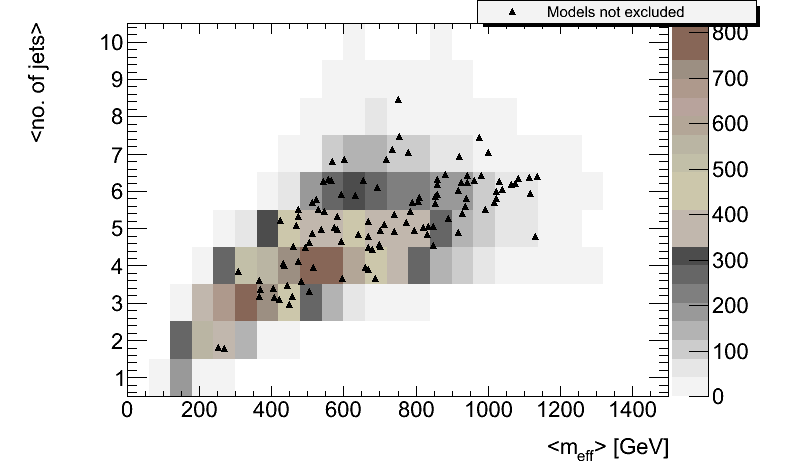}
   \caption{The distribution of  the average value of \meff\ to the average number of jets  for excluded points (colour scale) and not excluded points (black dots).}
   \label{fig:COLnjets-meff}
 \end{figure}

A lower cut on the effective mass, however, would cause a  significant increase in the
number of background events. We therefore studied additional features of these non-excluded models. 
A comprehensive study yields as the most significant 
feature a large average value of missing transverse momentum.
Figure \ref{fig:COLettom-meff} shows 
the ratio $f$ of the missing transverse momentum over the effective mass as a function 
of the effective mass. The not-excluded models at \meff\ $<600$~GeV do have average $f$-values above $0.3-0.35$. 
It is interesting to note that for higher \meff\ values smaller cuts on $f$ seem to be appropriate. 
Increasing the cuts on $f$
for the high \meff\ regions seems not to yield to an improved performance. 

In addition these points do show a typical 
average jet multiplicity as can be seen in Figure \ref{fig:COLnjets-meff}, 
also at \meff\ $<600$~GeV. The non-excluded points have jet multiplicities between $2-7$.

In conclusion we propose that ATLAS adds to future analyses signal 
regions with $f>0.3-0.35$ and a reduced effective mass
cut of \meff\ $>500$~GeV for high and low jet multiplicities.
A similar conclusion has been found for lower jet multiplicities in an independent study dedicated to 
compressed spectra \cite{LeCompte:2011fh}.
Some of the non-excluded points found in our study could 
be used as benchmark sets to further optimise the cut values with a detailed ATLAS simulation
including background events.

\subsection{Models from a logarithmic prior in the SUSY mass scale}

\begin{figure}[h]
  \centering
  \includegraphics[width=0.6\textwidth]{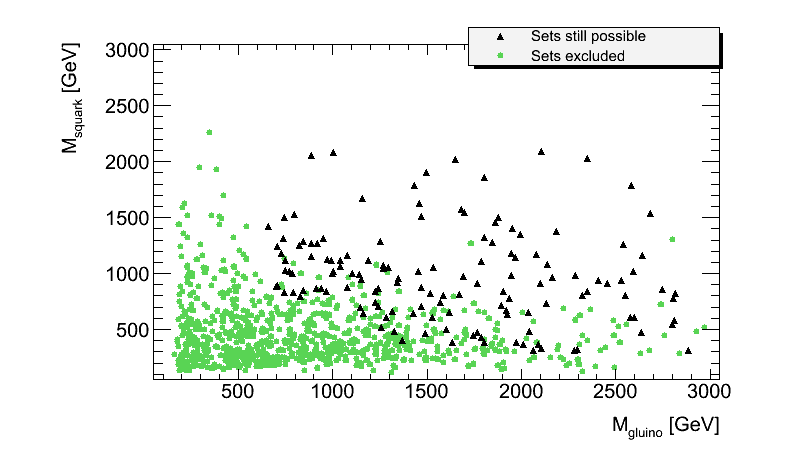}
  \caption{Exclusion range of the ATLAS analyses corresponding to \intlumi\ = 35 \lumipb\ and \intlumi\ = 1.04 \lumifb\ for 1000 randomly generated pMSSM points from a logarithmic prior; excluded model points are shown as green dots, not excluded models as black triangles.}
  \label{fig:exclM70}
\end{figure}

Figure~\ref{fig:exclM70} shows the result of our analysis of 
1000 points made with the logarithmic prior up to $3$~TeV 
in the mass scale of SUSY. 
Excluded points are shown as green points, not excluded ones as black triangles. 
Due to possible larger 
 masses of the squarks and gluinos more points survive at higher masses.
In total we find that 87\% of the model points are excluded.\\

\begin{figure}[h]
  \centering
  \includegraphics[width=0.6\textwidth]{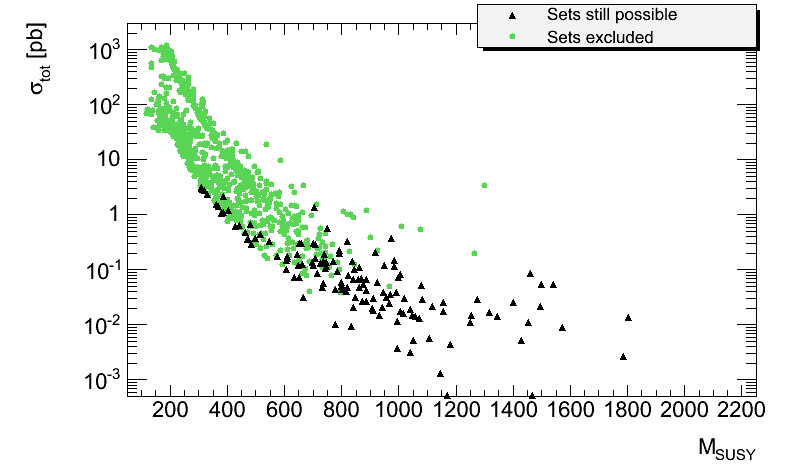}
  \caption{The total NLO squarks and gluino production cross section as a function of the 
minimal mass of the first and second generation squarks and the gluino $m_{\rm{SUSY}}$ for excluded 
model points (green dots) and not excluded models (black triangles). For some high mass 
model points the cross section is significantly enhanced by sbottom and stop production processes.}
  \label{fig:M70NLOmeff}
\end{figure} 

Again around one quarter of the not excluded model points 
have an average lepton number exceeding one. These points cannot be excluded because
 of the lepton veto.\\

A new feature is found in the logarithmic grid. Some SUSY models with gluino masses above 1000 GeV and squark masses 
between $300-600$~GeV are not excluded. Figure \ref{fig:M70NLOmeff} shows the total cross section for squark and gluino
production processes as a function of the SUSY mass scale $M_{\rm{SUSY}}$. All not-excluded 
SUSY models with
$M_{\rm{SUSY}}<600$~GeV are close to the {\it minimal} SUSY cross section 
at a given value of $M_{\rm{SUSY}}$.
The cross section is minimal since here only  
the \ensuremath{\tilde{d}_{\mathrm{R}}} and \ensuremath{\tilde{s}_{\mathrm{R}}} 
or 
the \ensuremath{\tilde{u}_{\mathrm{R}}} and \ensuremath{\tilde{c}_{\mathrm{R}}} 
are light. All other squarks and the gluino have much larger mass values.

These SUSY scenarios might also be missed in future searches, if 
the cuts on mass scale related variables (as \meff\ ) are raised further. 
Limits on squark masses derived at a minimum SUSY cross section might be helpful. 

In contrast to the flat-prior model points compressed mass 
spectra do not seem to be an important issue for the log-prior 
grid as far as we could infer from only 1000 model points.\\

\section{Summary}
We show that the \zerolepton\ of the ATLAS experiment excludes up to 99\% of the model points of 
the randomly generated pMSSM grid of ``Supersymmetry without Prejudice'' assuming a flat prior for a SUSY mass scale below 1 TeV. For 
the model points assuming a logarithmic prior up to 87\% are excluded.\\
Besides the models with a high average number of leptons, the most 
frequent reasons for the model points not to be excluded are a too
low cross section below the discovery potential, a
too low mass splitting between the lightest coloured sparticle and 
the neutralino resulting in a low effective mass \meff.
We propose to add selections with an increased missing transverse 
momentum cut and a decreased \meff\ cut, both with low and high jet multiplicities.
In addition we find that 
the search is quite insensitive if only one type right handed squarks 
is light, i.e. if the SUSY cross section is smaller than usually assumed.   
These scenarios might also profit from  
low mass signal regions with minimal statistical and systematic uncertainties. 

\section*{Acknowledgments}
We wish to thank Tom Rizzo, Carola Berger, James Gainer and JoAnne Hewett for providing the pMSSM model files. We thank Tom Rizzo for helpful comments.\\
We also thank Ben O'Leary for providing the cross section calculating program.

\appendix

\section{Additional tables}\label{app}

\begin{center}
\begin{scriptsize}
\begin{longtable}{cccccccccccc}
\caption{\label{tab:undiscprop}Important properties of some not excluded pMSSM models. Masses and energies are given in GeV, the cross section is given in $pb$.}\\ 
\toprule
\footnotesize model & \footnotesize m$_{\tilde{q}}$ & \footnotesize m$_{\tilde{g}}$ & \footnotesize m$_{\tilde{\chi}_0}$ & \footnotesize $\sigma_{NLO}$ & \footnotesize \etmis & \footnotesize \meff & \footnotesize N$_{Jets}$ & \footnotesize 1$^{st}$ p$^{\text{jet}}_{\text{T}}$ & \footnotesize 2$^{nd}$ p$^{\text{jet}}_{\text{T}}$  & \footnotesize N$_{Lep}$ & \footnotesize $\varDelta\phi$\\ 
\endhead
\toprule
81 & 741 & 913 & 442 & 0.31 & 232.7 & 692.0 & 5.8 & 241.2 & 139.1 & 0.5 & 1.2 \\ \midrule
155 & 676 & 963 & 232 & 0.39 & 262.1 & 847.6 & 5.1 & 296.1 & 187.2 & 1.0 & 1.2 \\ \midrule
175 & 629 & 841 & 502 & 0.46 & 230.4 & 667.8 & 4.7 & 239.4 & 132.4 & 0.3 & 1.2 \\ \midrule
247 & 805 & 890 & 310 & 0.23 & 262.4 & 861.0 & 6.4 & 297.8 & 188.3 & 1.6 & 1.1 \\ \midrule
638 & 930 & 999 & 145 & 0.1 & 351.9 & 1083.1 & 6.1 & 377.4 & 231.5 & 1.5 & 1.1 \\ \midrule
949 & 652 & 991 & 472 & 0.4 & 234.6 & 678.1 & 4.0 & 246.5 & 139.2 & 0.2 & 1.3 \\ \midrule
965 & 740 & 829 & 232 & 0.3 & 281.6 & 941.7 & 6.4 & 336.9 & 210.2 & 1.6 & 1.1 \\ \midrule
1294 & 880 & 782 & 416 & 0.29 & 222.0 & 719.1 & 7.0 & 234.6 & 153.6 & 1.0 & 1.1 \\ \midrule
1446 & 601 & 775 & 583 & 1.51 & 112.1 & 269.9 & 2.4 & 102.7 & 39.2 & 0.0 & 1.6 \\ \midrule
1956 & 684 & 821 & 127 & 0.44 & 261.6 & 999.4 & 6.5 & 379.6 & 235.9 & 2.2 & 1.0 \\ \midrule
2001 & 824 & 910 & 402 & 0.2 & 305.5 & 891.5 & 5.2 & 322.1 & 180.5 & 1.2 & 1.2 \\ \midrule
2041 & 566 & 908 & 469 & 1.09 & 159.5 & 449.5 & 3.5 & 164.0 & 88.5 & 0.4 & 1.4 \\ \midrule
2083 & 673 & 980 & 504 & 0.2 & 245.4 & 740.6 & 4.8 & 249.7 & 155.9 & 0.4 & 1.2 \\ \midrule
2157 & 809 & 872 & 473 & 0.19 & 247.3 & 733.2 & 6.6 & 251.1 & 143.8 & 1.0 & 1.1 \\ \midrule
2324 & 566 & 707 & 452 & 1.38 & 165.0 & 493.6 & 4.9 & 165.0 & 101.4 & 0.3 & 1.3 \\ \midrule
2409 & 553 & 866 & 485 & 1.51 & 132.7 & 366.2 & 3.3 & 132.5 & 68.7 & 0.2 & 1.4 \\ \midrule
2519 & 655 & 911 & 101 & 0.4 & 266.6 & 860.1 & 5.9 & 299.1 & 182.8 & 2.3 & 1.0 \\ \midrule
2577 & 839 & 944 & 434 & 0.2 & 283.1 & 882.4 & 6.1 & 312.3 & 188.4 & 0.7 & 1.1 \\ \midrule
2953 & 760 & 772 & 456 & 0.39 & 236.0 & 711.5 & 5.4 & 245.5 & 145.1 & 0.7 & 1.2 \\ \midrule
3226 & 827 & 646 & 404 & 0.73 & 179.8 & 602.3 & 6.9 & 212.4 & 126.3 & 0.7 & 1.0 \\ \midrule
3469 & 714 & 895 & 429 & 0.31 & 232.2 & 667.4 & 5.5 & 228.6 & 129.1 & 0.5 & 1.2 \\ \midrule
3666 & 782 & 787 & 520 & 0.38 & 183.3 & 565.1 & 6.4 & 192.3 & 114.7 & 0.7 & 1.1 \\ \midrule
3806 & 692 & 860 & 402 & 0.37 & 237.1 & 700.4 & 5.4 & 246.0 & 139.3 & 0.8 & 1.2 \\ \midrule
3809 & 724 & 811 & 604 & 0.39 & 173.9 & 503.3 & 4.7 & 171.6 & 100.8 & 0.6 & 1.3 \\ \midrule
4072 & 665 & 1026 & 483 & 0.29 & 206.6 & 583.6 & 4.8 & 209.8 & 109.0 & 2.1 & 1.2 \\ \midrule
4125 & 831 & 923 & 334 & 0.15 & 302.7 & 854.7 & 6.4 & 284.7 & 165.6 & 1.2 & 1.2 \\ \midrule
4293 & 741 & 968 & 167 & 0.23 & 298.9 & 961.9 & 5.7 & 336.2 & 210.2 & 2.1 & 1.1 \\ \midrule
4383 & 584 & 760 & 497 & 1.13 & 158.6 & 473.4 & 4.8 & 163.3 & 95.2 & 0.2 & 1.2 \\ \midrule
4570 & 800 & 703 & 509 & 0.61 & 194.1 & 593.6 & 5.6 & 201.7 & 120.9 & 0.3 & 1.1 \\ \midrule
5052 & 759 & 842 & 238 & 0.39 & 315.6 & 993.8 & 5.6 & 370.9 & 216.8 & 2.1 & 1.2 \\ \midrule
5312 & 558 & 765 & 514 & 1.02 & 163.5 & 433.5 & 3.7 & 156.7 & 75.1 & 0.6 & 1.3 \\ \midrule
5325 & 740 & 809 & 542 & 0.3 & 238.8 & 702.8 & 5.0 & 242.4 & 145.5 & 0.3 & 1.2 \\ \midrule
5360 & 737 & 938 & 262 & 0.39 & 263.5 & 835.3 & 5.3 & 285.5 & 181.4 & 1.4 & 1.2 \\ \midrule
5497 & 855 & 1005 & 207 & 0.11 & 349.6 & 1016.7 & 5.6 & 368.2 & 210.6 & 0.8 & 1.2 \\ \midrule
5700 & 623 & 899 & 555 & 0.71 & 148.0 & 405.4 & 3.5 & 146.9 & 74.9 & 0.1 & 1.4 \\ \midrule
5707 & 618 & 698 & 534 & 0.91 & 157.6 & 468.1 & 5.1 & 156.6 & 92.7 & 0.0 & 1.2 \\ \midrule
5805 & 699 & 916 & 164 & 0.45 & 275.5 & 860.8 & 5.8 & 307.1 & 180.6 & 1.6 & 1.1 \\ \midrule
6232 & 786 & 729 & 536 & 0.6 & 178.3 & 543.6 & 5.6 & 180.2 & 112.3 & 0.2 & 1.2 \\ \midrule
6333 & 825 & 730 & 456 & 0.49 & 177.8 & 568.0 & 7.3 & 203.6 & 112.7 & 0.9 & 1.0 \\ \midrule
6880 & 638 & 993 & 281 & 0.46 & 255.9 & 741.3 & 4.6 & 260.7 & 148.6 & 0.8 & 1.2 \\ \midrule
7105 & 632 & 900 & 616 & 0.53 & 160.4 & 408.0 & 2.8 & 150.5 & 69.0 & 0.1 & 1.5 \\ \midrule
7426 & 746 & 692 & 523 & 0.75 & 169.0 & 512.6 & 5.6 & 174.1 & 103.3 & 0.2 & 1.1 \\ \midrule
7514 & 661 & 871 & 493 & 0.52 & 205.0 & 596.3 & 4.4 & 212.0 & 120.8 & 0.3 & 1.3 \\ \midrule
7727 & 534 & 767 & 139 & 1.16 & 210.9 & 809.2 & 6.0 & 293.1 & 188.7 & 2.5 & 1.0 \\ \midrule
7736 & 846 & 895 & 227 & 0.19 & 303.9 & 921.0 & 6.5 & 309.8 & 190.1 & 1.4 & 1.1 \\ \midrule
7751 & 782 & 903 & 144 & 0.23 & 311.2 & 1041.0 & 6.0 & 381.1 & 231.4 & 1.7 & 1.1 \\ \midrule
7782 & 586 & 860 & 544 & 0.87 & 167.8 & 443.1 & 3.2 & 164.6 & 79.2 & 0.1 & 1.4 \\ \midrule
7888 & 631 & 688 & 574 & 1.68 & 114.1 & 308.4 & 3.5 & 106.5 & 57.7 & 0.1 & 1.4 \\ \midrule
7944 & 821 & 1023 & 356 & 0.16 & 308.3 & 980.4 & 5.6 & 357.0 & 217.7 & 1.2 & 1.1 \\ \midrule
8034 & 714 & 808 & 520 & 0.61 & 198.6 & 582.7 & 4.6 & 202.2 & 120.6 & 0.2 & 1.3 \\ \midrule
8396 & 831 & 977 & 499 & 0.17 & 279.5 & 819.1 & 5.0 & 296.3 & 170.3 & 0.4 & 1.2 \\ \midrule
8589 & 767 & 938 & 218 & 0.25 & 288.3 & 918.2 & 6.3 & 327.4 & 199.8 & 1.5 & 1.1 \\ \midrule
8686 & 572 & 770 & 554 & 0.87 & 164.4 & 421.8 & 3.1 & 153.4 & 71.5 & 0.1 & 1.5 \\ \midrule
8915 & 686 & 878 & 353 & 0.31 & 278.0 & 795.4 & 5.0 & 280.7 & 157.4 & 1.5 & 1.2 \\ \midrule
8916 & 784 & 895 & 343 & 0.19 & 270.7 & 852.8 & 6.0 & 280.0 & 183.8 & 1.3 & 1.1 \\ \midrule
9396 & 707 & 978 & 538 & 0.24 & 235.9 & 658.1 & 4.1 & 238.2 & 128.0 & 1.0 & 1.3 \\ \midrule
9497 & 583 & 960 & 435 & 1.18 & 167.2 & 482.9 & 3.9 & 174.6 & 96.5 & 0.5 & 1.3 \\ \midrule
9759 & 669 & 672 & 472 & 1.23 & 151.3 & 475.2 & 5.6 & 169.8 & 95.6 & 0.4 & 1.0 \\ \midrule
9781 & 563 & 856 & 507 & 1.3 & 134.9 & 368.5 & 3.4 & 132.7 & 67.1 & 0.4 & 1.4 \\ \midrule
10019 & 735 & 694 & 552 & 0.82 & 158.6 & 461.0 & 5.2 & 159.6 & 88.0 & 0.4 & 1.2 \\ \midrule
10149 & 768 & 789 & 530 & 0.41 & 216.7 & 639.1 & 4.9 & 221.3 & 129.9 & 0.4 & 1.2 \\ \midrule
10312 & 594 & 867 & 525 & 0.85 & 157.6 & 435.1 & 3.7 & 157.1 & 81.0 & 0.2 & 1.3 \\ \midrule
10531 & 624 & 892 & 610 & 1.01 & 107.9 & 253.5 & 2.2 & 97.4 & 35.4 & 0.0 & 1.6 \\ \midrule
10698 & 832 & 866 & 318 & 0.21 & 299.9 & 939.5 & 5.9 & 332.3 & 201.9 & 1.4 & 1.1 \\ \midrule
10923 & 841 & 659 & 542 & 0.74 & 173.9 & 512.1 & 5.3 & 180.3 & 98.8 & 0.3 & 1.1 \\ \midrule
11017 & 673 & 843 & 114 & 0.45 & 291.4 & 942.7 & 5.9 & 328.9 & 205.3 & 2.1 & 1.1 \\ \midrule
11050 & 662 & 1006 & 304 & 0.26 & 271.7 & 785.6 & 4.9 & 278.8 & 154.1 & 1.1 & 1.2 \\ \midrule
12020 & 583 & 830 & 495 & 1.09 & 180.3 & 503.7 & 3.4 & 182.4 & 100.5 & 0.0 & 1.4 \\ \midrule
12077 & 670 & 987 & 107 & 0.3 & 305.0 & 935.7 & 5.6 & 340.2 & 195.3 & 2.0 & 1.0 \\ \midrule
12942 & 763 & 986 & 424 & 0.2 & 250.5 & 789.8 & 6.1 & 281.0 & 167.6 & 0.5 & 1.1 \\ \midrule
13170 & 811 & 763 & 634 & 0.32 & 180.6 & 514.9 & 4.4 & 177.5 & 99.8 & 0.3 & 1.3 \\ \midrule
13187 & 622 & 822 & 437 & 0.68 & 192.0 & 593.8 & 5.5 & 208.8 & 122.5 & 0.4 & 1.1 \\ \midrule
13484 & 626 & 758 & 459 & 0.75 & 184.1 & 573.9 & 5.3 & 192.8 & 121.9 & 0.3 & 1.2 \\ \midrule
13616 & 647 & 970 & 445 & 0.48 & 229.2 & 667.4 & 3.9 & 236.7 & 140.7 & 0.1 & 1.4 \\ \midrule
13634 & 949 & 1006 & 395 & 0.09 & 336.8 & 1024.6 & 5.8 & 358.3 & 220.3 & 0.5 & 1.2 \\ \midrule
13698 & 721 & 971 & 246 & 0.4 & 276.3 & 925.8 & 5.5 & 326.4 & 205.7 & 1.7 & 1.1 \\ \midrule
13900 & 795 & 814 & 491 & 0.37 & 234.1 & 699.3 & 5.2 & 247.8 & 144.4 & 0.6 & 1.2 \\ \midrule
13938 & 566 & 800 & 508 & 0.64 & 190.5 & 546.7 & 4.5 & 194.1 & 104.3 & 0.7 & 1.2 \\ \midrule
14537 & 825 & 1026 & 273 & 0.14 & 311.8 & 918.1 & 5.4 & 327.5 & 184.0 & 1.0 & 1.2 \\ \midrule
14962 & 796 & 1016 & 125 & 0.18 & 320.8 & 1063.5 & 5.5 & 383.2 & 240.5 & 1.3 & 1.1 \\ \midrule
15788 & 786 & 894 & 240 & 0.34 & 257.1 & 807.2 & 6.0 & 286.5 & 171.9 & 1.2 & 1.1 \\ \midrule
15922 & 882 & 799 & 432 & 0.21 & 242.3 & 755.4 & 7.2 & 247.5 & 158.2 & 0.9 & 1.1 \\ \midrule
15940 & 859 & 1015 & 319 & 0.13 & 326.6 & 1030.4 & 5.8 & 372.8 & 226.7 & 0.7 & 1.1 \\ \midrule
15963 & 739 & 967 & 211 & 0.28 & 284.8 & 832.0 & 5.3 & 289.1 & 163.9 & 1.0 & 1.2 \\ \midrule
16092 & 682 & 760 & 454 & 0.59 & 212.8 & 654.0 & 5.7 & 224.3 & 136.0 & 0.5 & 1.2 \\ \midrule
16179 & 702 & 899 & 393 & 0.29 & 280.8 & 848.2 & 5.2 & 303.4 & 177.4 & 1.7 & 1.2 \\ \midrule
16379 & 812 & 988 & 109 & 0.14 & 333.3 & 1137.5 & 5.7 & 412.5 & 263.7 & 1.9 & 1.0 \\ \midrule
16411 & 754 & 632 & 498 & 0.97 & 182.7 & 539.7 & 5.2 & 190.6 & 103.8 & 0.4 & 1.1 \\ \midrule
16467 & 699 & 922 & 466 & 0.47 & 235.3 & 686.3 & 4.4 & 243.7 & 142.3 & 0.3 & 1.3 \\ \midrule
16539 & 733 & 1017 & 138 & 0.33 & 281.6 & 1022.6 & 5.8 & 382.3 & 241.5 & 2.0 & 1.0 \\ \midrule
16699 & 670 & 699 & 520 & 1.24 & 145.6 & 424.0 & 5.0 & 147.1 & 83.7 & 0.1 & 1.2 \\ \midrule
16722 & 590 & 945 & 505 & 0.59 & 172.2 & 474.5 & 3.9 & 174.2 & 86.9 & 0.2 & 1.3 \\ \midrule
17158 & 825 & 1003 & 741 & 0.19 & 131.5 & 365.5 & 3.6 & 126.6 & 72.0 & 0.2 & 1.4 \\ \midrule
17174 & 882 & 919 & 98 & 0.17 & 331.6 & 1117.8 & 6.4 & 410.0 & 253.1 & 1.1 & 1.1 \\ \midrule
17262 & 863 & 807 & 397 & 0.23 & 246.0 & 752.8 & 7.6 & 239.3 & 156.2 & 1.0 & 1.1 \\ \midrule
17559 & 820 & 712 & 548 & 0.53 & 173.7 & 531.0 & 5.7 & 178.3 & 107.6 & 0.4 & 1.1 \\ \midrule
17632 & 971 & 927 & 249 & 0.14 & 328.7 & 975.1 & 6.6 & 344.3 & 204.3 & 0.8 & 1.2 \\ \midrule
17767 & 659 & 844 & 492 & 0.4 & 182.6 & 556.8 & 6.1 & 197.5 & 108.9 & 0.5 & 1.1 \\ \midrule
18551 & 737 & 1010 & 190 & 0.19 & 316.5 & 1113.0 & 6.2 & 417.2 & 260.9 & 1.4 & 1.0 \\ \midrule
18564 & 1009 & 1018 & 234 & 0.07 & 413.2 & 1130.7 & 5.4 & 412.4 & 229.6 & 1.0 & 1.2 \\ \midrule
18884 & 802 & 1003 & 70 & 0.13 & 350.4 & 1076.5 & 6.1 & 400.0 & 233.0 & 1.0 & 1.1 \\ \midrule
18895 & 569 & 947 & 225 & 0.4 & 282.4 & 929.8 & 5.4 & 347.8 & 203.9 & 2.7 & 1.1 \\ \midrule
19016 & 776 & 706 & 538 & 0.61 & 171.4 & 523.7 & 5.8 & 174.9 & 106.4 & 0.3 & 1.1 \\ \midrule
19229 & 613 & 679 & 316 & 0.84 & 212.9 & 631.6 & 6.1 & 221.2 & 122.1 & 2.0 & 1.0 \\ \midrule
19367 & 696 & 1006 & 403 & 0.33 & 251.2 & 772.1 & 5.1 & 278.8 & 162.1 & 0.4 & 1.2 \\ \midrule
19707 & 789 & 845 & 340 & 0.26 & 251.8 & 779.2 & 6.6 & 264.2 & 160.6 & 0.8 & 1.1 \\ \midrule
19740 & 578 & 1011 & 480 & 0.96 & 163.1 & 457.4 & 3.2 & 169.6 & 91.3 & 0.0 & 1.4 \\ \midrule
19750 & 731 & 785 & 475 & 0.5 & 223.0 & 669.5 & 5.0 & 233.2 & 139.8 & 0.4 & 1.2 \\
\bottomrule 
\end{longtable}
\end{scriptsize}
\end{center}

\end{document}